# scientific reports

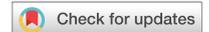

OPEN

# Correlation between active regions' spectra at high radio frequencies and solar flare occurrences

Sara Mulas[1,11✉], Alberto Pellizzoni[1,11], Marco Marongiu[1,11], Adriana Marcucci[2,11], Simona Righini[3], Maria Noemi Iacolina[4], Elise Egron[1], Giulia Murtas[5], Matteo Bachetti[1], Francesco Berrilli[6], Alessandro Cabras[1], Roberto Caocci[1], Gian Luigi Deiana[1], Salvatore Luigi Guglielmino[7], Colby Haggerty[5], Adelaide Ladu[1], Sara Loru[1], Andrea Maccaferri[3], Pasqualino Marongiu[1], Andrea Melis[1], Alessandro Navarrini[1], Alessandro Orfei[3], Pierluigi Ortu[1], Mauro Pili[1], Tonino Pisanu[1], Giuseppe Pupillo[3], Andrea Saba[4], Luca Schirru[1], Giampaolo Serra[4], Caterina Tiburzi[1], Giuseppe Valente[4], Alessandra Zanichelli[3], Pietro Zucca[8] & Mauro Messerotti[9,10]

High radio frequencies observations with the Italian network of large single-dish radio telescopes resulted in ~450 solar images between 2018 and 2023 in K-band frequency range (18–26 GHz). Solar radio mapping at these frequencies allows the probing of the Active Regions (ARs) chromospheric magnetic field close to the Transition Region, where strong flares and coronal mass ejection events occur. Enhanced magnetic fields up to 1500–2000 G determine anomalous spectra in the ARs brightness compared to pure free-free emission, due to the addition of a steeper gyro-resonance component also associated with circular polarisation up to ~40%. When a significant AR spectral flattening is detected, the probability of a strong flare occurrence within ~30 hours is high (~89% in terms of statistical precision). Despite an approximate weekly cadence of our observations, only ~12% of strong flares are missed/unpredicted within this time interval. Through a correlation analysis, we assess the trade-off on the sensitivity and the robustness of this physics-based flare forecast method.

Mapping the brightness temperature of the solar atmosphere at high radio frequency (>1 GHz) reveals incoherent free-free emission originating primarily from plasma processes in the local thermodynamic equilibrium, with the addition of sporadic gyro-magnetic emission. These relatively simple processes, compared to other frequencies, provide a probe of physical conditions and vertical structure of the Chromosphere up to the Transition Region and the Corona[1]. The contribution to the Active Regions (ARs) brightness temperature is at least partially connected to gyro-resonance emission at frequencies <50 GHz, while thermal bremsstrahlung is the only emission component seen at high frequencies 100–400 GHz[2,3] show that the cm-millimetric emission (17 and 34 GHz) in the total flux density spectra of ARs systematically exceeds the expected fluxes from the optically thick free-free emission. This radio emission is due to the concurring presence of (1) gyro-resonance emission at 17 GHz[4–6], (2) optically thick thermal free-free emission from the upper Chromosphere, and (3) optically thin thermal emission from the Transition Region and coronal sources. A 3D solar atmospheric model, developed to reproduce the brightness temperature of a specific AR characterized by a high degree of polarisation (85%) with radio observations at 17 and 34 GHz from NoRH[7], showed that at 17 GHz the emission was successfully modelled as gyro-resonance, while at 34 GHz the emission was attributed to free-free radiation only.

[1]Cagliari Astronomical Observatory, INAF, Via della Scienza 5, 09047 Selargius (CA), Italy. [2]CNMCA, ITAF, Via Pratica di Mare 45, 00040 Pomezia (RM), Italy. [3]Institute of Radio Astronomy, INAF, Via Gobetti 101, 40129 Bologna, Italy. [4]c/o Cagliari Astronomical Observatory, ASI, Via della Scienza 5, 09047 Selargius (CA), Italy. [5]University of Hawai'i at Mānoa, Street, 96822 Honolulu, USA. [6]Department of Physics, University of Rome Tor Vergata, Via della Ricerca Scientifica 1, 00133 Rome, Italy. [7]Catania Astrophysical Observatory, INAF, Via Santa Sofia 78, 95123 Catania, Italy. [8]The Netherlands Institute for Radio Astronomy, ASTRON, Oude Hoogeveensedijk 4, 7991 Dwingeloo, The Netherlands. [9]Trieste Astronomical Observatory, INAF, Via Giambattista Tiepolo 11, 34131 Trieste, Italy. [10]Department of Physics, University of Trieste, Via Alfonso Valerio 2, 34127 Trieste, Italy. [11]These authors contributed equally to this work: Sara Mulas, Alberto Pellizzoni, Marco Marongiu and Adriana Marcucci. ✉email: sara.mulas@inaf.it

From 2018 to 2023, the radio imaging system of the large single-dish telescopes of the Italian National Institute for Astrophysics (INAF https://www.radiotelescopes.inaf.it) has produced ∼450 full solar disk images in the K-band (18–26 GHz) frequency range (SunDish project https://sites.google.com/inaf.it/sundish). This was possible thanks to an instrumental upgrade of the system suited for solar observations[8]. A first development phase aimed at defining and optimising solar observing configurations, developing techniques for the radio telescope imaging process, and sorting calibration issues[8–11]. We then provided early scientific results, including a catalogue of detected ARs, an accurate measurement of the brightness temperature of the radio quiet-Sun (QS) component, the solar radius, the outer coronal emission, and preliminary constraints to atmospheric models in terms of density and temperature distributions[12–14]. We also showed evidence for peculiar ARs spectral variations, related to variable gyro-resonance emission components that anticipate flare events[8].

The changes in ARs gyro-resonance spectral components in K-band result from the evolution in topology of the magnetic fields at the boundary between the Chromosphere and the Transition Region (e.g.[15]). Therefore, these spectral variations can be treated as a valuable proxy for eruptive plasma processes occurring in the ARs. Such signatures are worthy of further investigations, since developing an approach for predicting solar flares is essential to mitigate possible storm-like geo-effectiveness and related hazards (e.g.[16]), a core research in the perspective of operational Space Weather research. Despite extensive worldwide efforts, a robust, accurate, and sensitive operational tool for flare forecasting is still missing[17–19]. Several works have been conducted to identify a reliable method for flare forecasting. Some researchers designed statistical models to predict the occurrence of flares, based on the physical properties of ARs[20–22]. Others employed machine learning methods for flare forecasting, based on the availability of large amounts of flare-related data[23–25]. An example is FLARECAST (Flare Likelihood And Region Eruption Forecasting), a flare prediction system based on the automatic extraction of physical AR properties such as the area, magnetic flux, shear, magnetic complexity, helicity and proxies for magnetic energy extracted from solar magnetogram and white-light images[26]. In alternative to these complex machine-learning approaches[27], derived physics-based thresholds for the onset of major solar flares, predicting them within the next 20 hours through a critical condition of magnetohydrodynamic instability. Our approach complements these methods, as it is specifically focused on probing the solar atmospheric layers where most of the flare phenomenology is expected to occur, i.e. the upper Chromosphere near the Transition Region, through high-frequency radio observations[8,28].

## Results

We reprocessed and updated the ARs spectral index ($\alpha_{T_p}$) distribution obtained from our K-band images[8], by comparing ARs peak brightness temperatures ($T_p$) at different frequencies between 2018 and 2023 (Fig. 1). The spectral index values are calculated as follows:

$$\alpha_{T_p} = \frac{log[(T_{p1}/T_{p2})(\nu_1/\nu_2)^2]}{log(\nu_1/\nu_2)} \ .$$

(1)

where $T_{p1}$ and $T_{p2}$ are the ARs peak brightness temperatures at the frequencies $\nu_1$ and $\nu_2$[3].

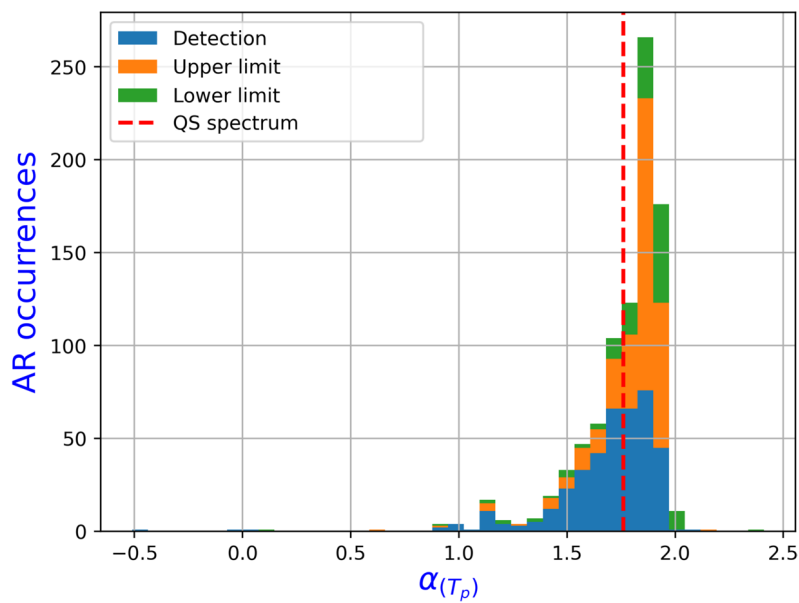

**Fig. 1.** Histogram of the spectral index values calculated from the maximum brightness temperatures $T_p$ of the ARs in the 18–26 GHz frequency range for the period 2018–2023. Blue counts indicate spectral index measurements when the AR is detected at both frequencies, orange and green counts show upper limits and lower limits, respectively. The dashed line indicates the spectral index of the QS-level[8]. Spectral index errors range between 0.02 and 0.36, with an average value of 0.05.

The distribution peaks at the expected spectral index (Rayleigh-Jeans emission with an $\alpha_{RJ} \sim 1.9$) for thermal bremsstrahlung in the optically thick regime, as observed in ARs at higher frequencies[2,3,29]. On the other hand, Fig. 1 shows that $\sim 12\%$ of the ARs spectra (also including upper limits) are significantly softer ($\alpha_{T_p} < 1.5$) than the dominant free-free emission, confirming important clues about AR spectra. The uncertainties $\sigma_\alpha$ on the derived spectral indexes are typically <0.1 (see[8]), and hence $\alpha_{T_p} < 1.5$ results inconsistent with thermal Rayleigh-Jeans (RJ) emission at $>4\sigma_\alpha$.

Fig. 2 highlights the typical differences between ARs with a RJ and an anomalous spectrum. AR SPoCA 24694 observed on the 29th of September 2020 with the Medicina Radio Telescope, appears to be very bright both in the 18 GHz (Fig. 2a) and in the 26 GHz images (Fig. 2b) due to the hard spectral index (1.75). On the other hand, in the circular polarisation percentage images (Fig. 2c) the same AR is barely visible, with a peak value of $\sim 1\%$.

The lack of a significantly bright AR flux at 26 GHz is a typical symptom of an anomalous spectrum, shaped by gyro-magnetic emission. A peculiar example is AR NOAA 12785 (located on the limb bottom left corner in Fig. 2d-e-f), detected on the 23rd of November 2020 with the Medicina Radio Telescope. The spectral index derived for this AR ($\alpha_{T_p}$ =1.15) corresponds to one of the lowest recorded in[8]. We stress that this AR presents a higher percentage of circular polarisation, $\sim 4\%$ (see Fig. 2f) compared to the ARs dominated by an RJ emission component. The association between spectral anomalies and a high circular polarisation level is coherent with the presence of strong magnetic fields, leading to gyro-resonance emission processes. After 12 hours from our observation, a C4.4 class flare erupted from the same AR. Other examples of associations between peculiar ARs and flare occurrences are extensively discussed in Supplementary Material Sect. 1.

These anomalous AR spectral indexes can be related to emerging gyro-resonance emission components due to enhanced magnetic fields[8,15]. We then probed their association with flare occurrences through a correlation analysis. To this scope, we defined a "strong flare" threshold based on the flare class and the solar cycle phase (see Methods for more details). When considering the association between very flat and polarised ARs spectra and strong flares, we noticed it is not uncommon to see a delay up to 1–2 days between the two. Such large delays are also reported in the literature related to precursors of flares based on peculiar ARs magnetic configurations (see[27]). Taking these factors into consideration, we chose 30 hours as the maximum temporal interval limit required to consider a positive association between AR anomalous spectral detection ($\alpha_{T_p} < 1.5$) and flare occurrence. Such a time threshold was chosen after a careful optimization analysis, as discussed in Sect. Methods.

Table 2 contains a comprehensive list of all the ARs spectra used in the correlation analysis and their association with flare occurrences. The general result of the correlation analysis shows that events with positive anomalous spectral index-flare association (True Positive; TP) are $\sim 80\%$ of the total entries in the Table 2, while significant spectral flattening not associated with strong flare occurrences (False Positive; FP), and intense flare without anomalous spectrum association (False Negative; FN) are both $\sim 10\%$. The larger remaining class of True Negative (TN) is not providing significant additional information suitable for Table 2. Table 1 presents the confusion matrix related to the performances of adopting anomalous AR spectral indexes as flare proxies.

When an anomalous spectral index is detected there is a $\sim 89\%$ probability (precision: $\frac{TP}{TP+FP}$) of an intense flare occurring within $\sim 30$ hours. On the other hand, an $\alpha_{T_p}$ >1.5 is a TN in 97% of cases (specificity: $\frac{TN}{TN+FP}$). Only $\sim 12\%$ of the intense flares are missed by this forecast method (miss rate: $\frac{FN}{TP+FN}$), although most of them still present AR spectral index close to our $\alpha_{T_p}$ =1.5 threshold. The sensitivity ($\frac{TP}{TP+FN}$) is $\sim 88\%$, while the false alarm ratio ($\frac{FP}{TP+FP}$) is ∼11%. Since we have an unbalanced data set, we also calculate the the harmonic mean of the model's precision and sensitivity or F1-score ($\frac{2TP}{2TP+FP+FN}$), which is ∼89%.

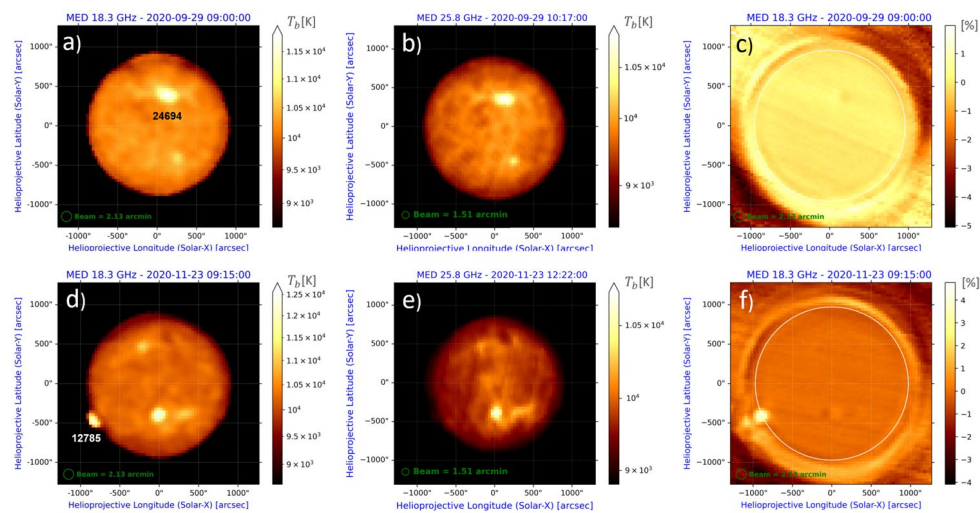

**Fig. 2.** Sample of solar images showing ARs spectra and circular polarisation typically associated to Rayleigh-Jeans (first row) and gyro-resonance emission (second row) at two different epochs, acquired with the "Grueff" Medicina Radio Telescope. Panels a and d represent the Total Intensity in brightness temperature ($T_b$) at low K-band frequencies ($\sim 18$ GHz); panels b and e in the higher K-band ($\sim 26$ GHz). Panels c and f depict the circular polarisation percentage (LHCP is positive) at low K-band ($\sim 18$ GHz).

|  | Predicted positive | Predicted negative |
|---|---|---|
| **Actual Positive** | TP = 99 | FN = 13 |
| **Actual Negative** | FP = 12 | TN = 434 |

**Table 1**. Flare forecasting method performance evaluation: Confusion Matrix showing total number of cases for each category (True Positive, False Negative, False Positive and True Negative.).

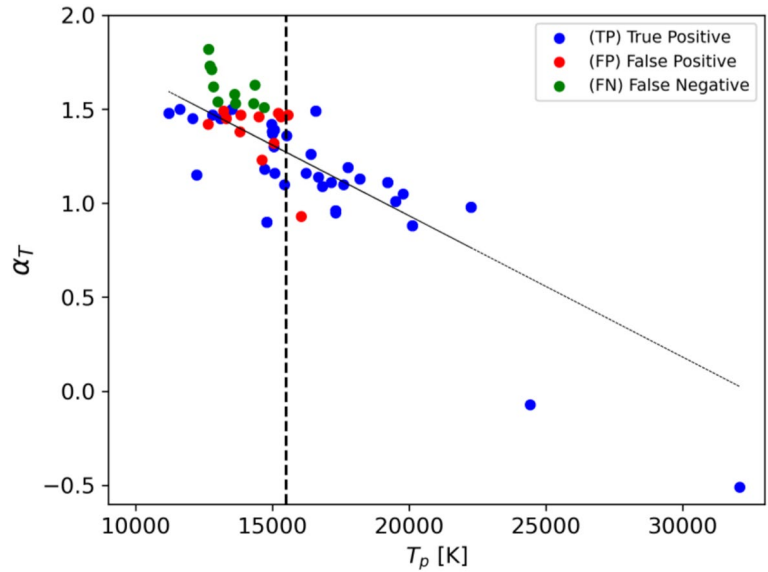

**Fig. 3**. Spectral index as a function of the ARs peak brightness temperature $T_p$. Blue, red and green dots indicate the three categories TP, FP and FN respectively. The dashed vertical line ($T_p = 15500K$) separates ARs with peak brightness temperature above ∼50% the QS level.

Close to the solar disk limb, the AR spectral estimation can be affected by observational biases such as the lack of visibility of the whole AR. We define an AR to be on the limb if its distance from the centre of the Sun exceeds 95% of the solar disk radius (see[9] for more details). If we exclude the 24 AR-flare limb events from our statistics, both the precision and the F1-score increase only slightly at the expense of the inability to predict flares close to the limb, with no significant improvement in our prediction method.

These performances are based only on the usage of spectral AR information. The inclusion of other relevant AR parameters in the flare prediction method must also be explored.

As anomalous AR spectra are associated with gyro-resonance components that add up to the background thermal emission, it is worthwhile to examine the effect of the anomalous brightness. In Fig. 3 we display the relation between the AR spectral index and $T_p$, showing an anti-correlation between the peak brightness temperature and the spectral index (the Pearson correlation coefficient is $C_t = -0.74$). The observed phenomenon is also confirmed in a work by[30], with data from the Metsähovi Radio Observatory (MRO) of Aalto University, who recently stated that intense maximum radio brightening is needed in ARs to cause a solar flare.

We tested the inclusion of an AR brightness threshold in our flare prediction method: the TP cases would be defined as ARs with anomalous spectrum ($\alpha_{T_p} < 1.5$) and a peak brightness exceeding ∼50% the QS level (>15500 K). From Fig. 4 we can see that most of the FP and FN events are associated with $T_p$ lower than 15500 K. Therefore, by combining this additional condition on the $T_p$ value, we are able to better filter out these cases, leading to an increase of the precision of the confusion matrix (from ∼ 89% to ∼ 97%). On the other hand, there are several TP cases with an excess lower than 5000 K. As a consequence, the flare miss rate rises significantly (from ∼ 12% to ∼ 40%). If we consider flares from M and X classes only, the correlation coefficient between $\alpha_{T_p}$ and $T_p$ rises to $C_t$=−0.8, although the statistics are less precise and the number of events is much lower.

Additional investigation on polarimetric mapping of solar disk emission features could allow us to correlate magnetic information with AR spectra, and to disentangle gyro-magnetic contributions from free-free emission. The inclusion of Stokes V in a more complex detection scheme could be considered, although the circular polarisation intensity is also related to geometrical and line-of-sight issues. The "Grueff" Medicina Radio Telescope permits dual-polarisation observing mode, and we challenged such information for solar imaging. By combining the Stokes parameters I and V from our maps, we obtained the fractional polarisation $\rho = \frac{StokesV}{StokesI}$, where Stokes I=$\frac{1}{2}(LHCP + RHCP)$ (see[31]) and converted it in a percentage value ($\rho_{\%}$). We adopted the convention Stokes V=$\frac{1}{2}(LHCP - RHCP)$, therefore negative values of $\rho_\%$ are associated with RHCP while positive values with LHCP. Most of the disk exhibits a QS component almost circularly unpolarised, with an absolute value of $\rho_\%$ <0.1%. The majority of the flaring ARs (∼80%) have instead a polarisation percentage larger than 1%. In principle, higher values of polarisation degree in AR can be explained by a stronger magnetic field in sunspots than in the QS. In Fig. 5 we divided the events into three categories (TP, FP and FN) according to the value of $\rho_\%$ in order to analyse the impact of this further parameter on our flare prediction method. Compared to the $T_p$, the $\rho_\%$ fits better the TP cases, however it is less helpful in filtering the FP and FN cases. While $\alpha_{T_p}$ does not depend on the line of sight, the polarisation percentage is heavily influenced by the geometry. This could explain why ∼ 20% of the TP cases have a $\rho_\%$ lower than 1%. In the Supplementary Material Sect. 1 and Sect. 2, we also discuss how the correlation between Hale classification of the ARs photospheric magnetic complexity[32,33] and flare occurrences[34–36] could be coupled with $\rho_\%$ information from our images. Furthermore, in the Supplementary Material Sect. 3 we investigated a possible correlation among different AR physical parameters such as the time delay $\Delta t$, the degree of AR spectral steepening, the flare and polarisation intensity.

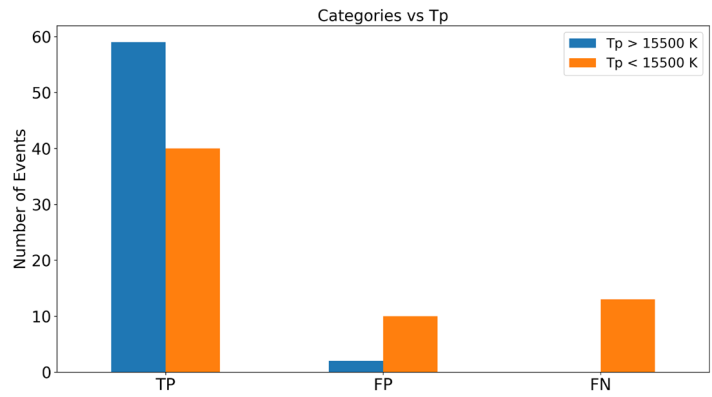

**Fig. 4.** Histogram of the number of ARs with a maximum brightness temperature ($T_p$) higher (blue) and lower (orange) than 15500 K for the three categories TP, FN and FP.

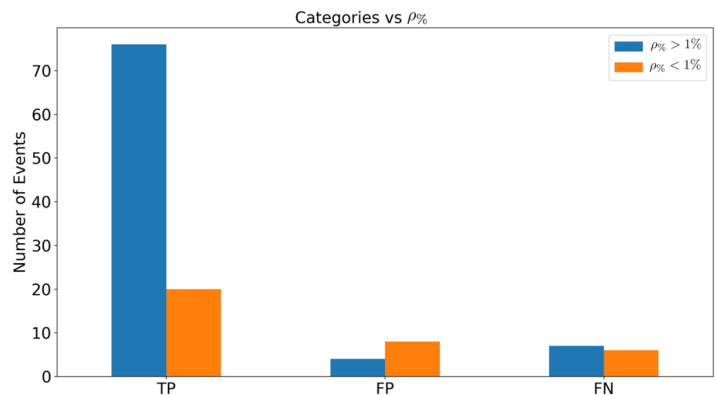

**Fig. 5.** Histogram of the number of ARs with circular polarised emission higher (blue) and lower (orange) than 1% (absolute values) for the three categories TP, FN and FP.

Thanks to the $\rho_\%$ parameter obtained from our data, we can estimate the magnitude of the chromospheric longitudinal magnetic field above ARs. Under the assumption of gyro-resonance source emission, the following numerical relation holds between the observing frequency and the magnetic field:

$$B[\text{G}] = \frac{10700}{s \cdot \lambda[\text{cm}]} = \frac{360 \cdot f[\text{GHz}]}{s} \qquad (2)$$

where *s* is an integer number related to the harmonic of the gyro-resonance component and *f* is the observing frequency. Gyro-resonance emission is produced by thermal electrons gyrating around the magnetic field lines, and it is strong in regions where the observing frequency is a low order harmonic of the electron gyro frequency[31]. For example, we consider the strong polarised emission (>8%) from AR 12786 ($\alpha_{T_p} = 1.37$) at 18.3 GHz (Supplementary Material Fig. S3), where a cluster of C-class flares was detected within 24 hours. Assuming gyro-resonance emission, the magnetic field above AR 12786 estimated according to Eq. 2 is between ∼ 2100$G$, in case of the third harmonic (s=3), and ∼ 3200$G$ for the second harmonic (s=2). Higher order harmonics were not considered since s=2,3 are the most commonly observed, because they are emitted more efficiently and suffer less absorption. Thus, enhanced magnetic fields (up to 1500–2000 G) in the upper Chromosphere close to the Transition Region determine anomalous spectra in the AR brightness temperature compared to pure free-free emission, due to the addition of a steeper gyro-resonance component also associated with circular polarisation up to ~40% (see Supplementary Material Sect. 1.1).

## Discussion

The gyro-resonance emission at radio frequencies has been studied by many authors in order to estimate the Transition Region and coronal magnetic field strength up to the 15–17 GHz range (e.g.[2,3,37–43]). They showed that the gyro-resonance emission at radio frequencies >10 GHz is produced in regions with magnetic field intensities around the kG located close to the Transition Region. At 17 GHz, the lower limit for the magnetic field intensity to produce this emission (due to the third harmonic) is ~2000 G at the photospheric level. They also estimated a magnetic field to be at least of 2700 G at the base of the Transition Region, and 1800 G in the low corona. Moreover, the enhancement of gyro-magnetic effects when leaving the minimum of solar activity can cause a significant increase in the AR average brightness up to $10^5$ K and higher. The gyro-resonance emission was proven to peak up to ~20 GHz (see, e.g.[7]), although in most cases, this component is observed at lower frequencies (2–5 GHz[44,45]). We corroborated this scenario in our early results claiming sporadic gyro-magnetic emission up to 26 GHz[8] and confirmed that the concurring presence of sporadic gyro-magnetic components in the AR emission could contribute to the spectral flattening.

In the present work we confirmed that peculiar spectra ($\alpha_{Tp} \leq 1.5$) affecting ~12% of our ARs sample are significantly flatter than RJ emission, and can be related to the contribution of a steep high-frequency tail provided by gyro-resonance emission peaking slightly below our frequency range. The presence of gyro-resonance emission components in the flat-spectrum ARs is also corroborated by their associated significant circular polarisation up to ~40%.

A possible physical context for these spectral outliers is connected with local eruptive events inside ARs – including flares and coronal mass ejections (CMEs) – which are crucial to investigate the AR dynamics. A gyro-magnetic emission enhancement could be a precursor of flaring gyro-synchrotron emission from the region of anomalous energy release with a sub-relativistic plasma in moderate magnetic fields[46], and modelled according to several approaches (e.g.[47,48]). Broadband analysis of these eruptive events shows changes – a few days before the event – in their microwave/radio spectra, suggesting both the emergence of a new AR photospheric magnetic field and shifting movements of the sunspots[1,49–51]. These changes can be used as a predictive criterion for eruptive events on the Sun. In particular, several papers (e.g.[50,51]) show a correlation between changes in the ARs magnetic field configuration taking place one or two days prior in the same region and the eruptive events. This could lead to a change in the balance between the thermal and non-thermal emission components, with a prevalence of the gyro-magnetic emission that produces an inversion from positive to negative values in the spectral index.

In fact, the adoption of the AR radio spectral index in K-band as a proxy, specifically related to the dynamic of upper solar chromospheric layers, shows robust performances as a flare forecasting method (characterised by ~89% precision and ~12% miss rate). By adding ARs radio brightness and magnetic field information, the precision improves up to 97%. Therefore, they are precious parameters to filter the FP and FN cases, however, they significantly increase the miss rate. On the other hand, present limitations of our method are related to the inability to specifically disentangle eruptive flares (i.e. associated to CMEs). Also, the time needed for the imaging observation procedure (2–3 hours) hinder the possibility of obtaining nowcasting parameters and continuously monitor the ARs during the flares.

The opportunity of regular monitoring of the solar Chromosphere, Transition Region and Corona at different radio frequencies could better constrain the spectral status of the ARs before, during and after flare events. Simultaneous higher frequency observations (e.g. MRO images at 34 GHz[30],) can be useful to precisely constrain the free-free emission component of the ARs, while the variable gyro-resonance emission can be probed at lower frequencies being sporadically the dominant process at 1–15 GHz[15].

The proposed flare prediction method, based only on radio imaging data, could be integrated in wider multi-messenger Space Weather prediction systems – e.g. including UV and Hα data probing the dynamic evolution of the ARs in the Chromosphere. This could offer a precious disentanglement among eruptive and non-eruptive flare events, as well as extended information about the vertical structure of the AR magnetic field, anticipating flaring events.

## Methods

In[8] we presented a catalogue of radio continuum images in the frequency range 18–26 GHz (~170 solar maps) including a multi-wavelength identification of ARs, their brightness and spectral characterization for the 2018–2020 period. We also described image processing, calibration and data analysis procedures related to solar observation modes implemented for the Sardinia Radio Telescope and the "Grueff" Medicina Radio Telescope, which are at the base of the present study. Further details on the image calibration and data analysis process are given in[11] and[9,10], respectively. In this work, we report the analysis of ~450 solar maps acquired between 2018 and 2023, that significantly increase the previous sample. We also adopted an improved version of the SUNPIT pipeline[10], suited for the data analysis and the imaging procedure of the solar maps for the SunDish project. This version allows us to derive circular polarisation (Stokes V and $\rho_{V\%}$ parameters) from both observations using spectro-polarimetric and total power backends (since left-/right-handed circular polarisation channels are available in both cases), a fundamental tool to investigate gyro-resonance processes.

In order to study the statistical correlation between flares occurrences and the characteristics of the ARs that produced them, we first identified all of the ARs associated with peculiarly low spectral indices ($\alpha_{T_p} < 1.5$) in the 2018-2023 archive. We classified 558 AR events for which spectral measurements including suitable lower/ upper limits were available (only upper limits with $\alpha_{T_p} < 1.5$ and lower limits with $\alpha_{T_p} > 1.5$ were included). Subsequently, an investigation was conducted to ascertain whether these phenomena were concomitant with intense flares occurring temporally close to the SunDish observations. This approach required the selection of (1) a suitable complete flare catalogue, (2) a flare class threshold arbitrarily defining the "strong flare" category, according to the flare statistics of the considered solar cycle phase, and (3) a time delay limit between AR anomalous spectral detection and flare occurrence. These parameters are discussed below, and trimmed to optimize the performance of a flare forecast method based on AR spectral features.

Concerning the flare catalogue, we considered all the events in[52], whose identification is based upon soft X-ray data (1–8 Å) by the NOAA/NASA GOES satellite constellation (https://www.ngdc.noaa.gov/stp/satellite/goes/dataaccess.html). Their catalogue has been recently extended up to April 2023[53]. For the remaining seven months of 2023, the presence of flares above our established thresholds for each anomalous AR was analysed directly from GOES data. We also adopted the flare catalogue by[54] to cross-check our results and fill any missing epochs in GOES data. This catalogue contains the solar flares detected by the Astro-rivelatore Gamma a Immagini LEggero (AGILE) satellite (https://www.asi.it/en/planets-stars-universe/alte-energie/agile/).

Since the 2018-2023 period covers both the solar minimum and the rising phase of the solar cycle 25, the flare rate and class distribution strongly evolve with time. From the previously mentioned catalogues we selected only a fraction (3%) of the strongest flares for each semester to be included in our correlation analysis. This implicitly defines a variable threshold for the minimum flare class depending on the period, as shown in Table 3.

When considering the association between very flat and polarised AR spectra and strong flares, we noticed it is not uncommon to see a delay up to 1–2 days between the two (see. e.g. the case of AR13477 on 60258 MJD or AR13220 on 59587 MJD from Table 2). Such large delays are also reported in the literature about precursors of flares based on peculiar ARs magnetic configurations (see[27]). Taking these factors into consideration, we chose 30 hours as the maximum temporal interval limit required to consider a positive association between ARs anomalous spectral detection and flare occurrence. We selected this time threshold after a careful optimization analysis, which we discuss later in the article.

The criteria described above have been used to divide our ARs spectral measurements into four categories, thus defining the rules for a flare prediction strategy:

- true positive cases (TP) where an AR with $\alpha_{T_p} < 1.5$ has been identified and a flare above our threshold intensity class coming from the same AR has been detected in the previous or following 30 hours from our observation;
- false positive cases (FP) where an AR with $\alpha_{T_p} < 1.5$ has been identified, but no flare above our threshold intensity class coming from the same AR has been detected within 30 hours following or preceding our observation;
- false negative cases (FN) where a flare above our threshold intensity class has been detected within the 30 hours preceding or following our observation but $\alpha_{T_p}$ of the associate AR is >1.5;
- true negative cases (TN) where $\alpha_{T_p} > 1.5$ and no flares above our threshold intensity class have been detected within 30 hours following or preceding our observation.

We explored possible improvements in the statistical performances by using different values for the spectral index threshold, the time delay ($\Delta t$) between the flare and the observation, and the flare class threshold.

The value 1.5 was chosen for the spectral index threshold in order to identify the ARs with a strong gyro-resonance contribution. A higher threshold value determines the inclusion of more ARs with a relevant RJ component, which leads to a significant increase of the FP without a particular change in the TP cases. As a consequence, the precision declines while the false alarm rate increases. We did not consider a lower threshold value, since it exceeds the limit of $4\sigma$ error from the expected value for a thermal free-free spectrum ($\alpha_{T_p} = 1.9 \pm 0.1$), and the TP value quickly decreases in this direction.

For the time delay parameter $\Delta t$, we initially chose 24 hours, but exploring longer time windows, we found that a $\Delta t$ of 30 hours allowed us to maximize the statistical performances. Although the number of FN increases slightly, the FP decreases drastically and the TP rises up to 20% compared to the 24 hours window. On the other hand, when extending the time window even longer, the proposed forecasting method sensitivity decreases without any significant improvement in the precision.

Almost all the detected ARs with $\alpha_{T_p} < 1$ are associated with M and X-class flares. There are only a few exceptions, such as AR13182 ($\alpha$=−0.07) on 59960 MJD that hosted a C8.6 flare. When our analysis is limited to solar flares belonging to M and X classes, the sensitivity of the proposed method rises from 89% to 94%. However, the precision decreases from 88% to 62%, because many ARs associated with C-class flares would be treated as FP cases. Therefore, we decided to not limit our statistical analysis to only the M and X-class flares to have the best trade off between the sensitivity and the precision.

We provide a comprehensive list of all the ARs spectra used in the statistical analysis presented in this paper (Table 2). The ARs are listed in chronological order according to the Modified Julian Date (MJD) of the observations. This table includes also the following parameters: the time distance between the map acquisition and the detection of any associated flare ($\Delta_T$) within 30 hours, the flare intensity (Flare Class), the spectral index of the AR ($\alpha_{T_p}$), the NOAA AR identification number ($AR_{ID}$), the category (Flag) of the event with respect to the solar flare occurrence, the maximum brightness temperature corresponding to the AR detection at 18 GHz ($T_p$), the magnetic complexity, and the $\rho_\%$ parameter obtained from polarisation maps. We calculated the uncertainty associated with the $\rho_\%$ as the RMS of a background region where no sunspots have been identified

| Epoch | $\Delta_T$ | Flare class | $\alpha_{T_P}$ | $AR_{ID}$ | Flag | $T_P$ | Magnetic complexity | $\rho\%$ |
|---|---|---|---|---|---|---|---|---|
| [MJD] | [h] | | | | | [K] | | [%] |
| 59151 | −12 +2 +10 | C1.5 C4.3 C1.6 | 1.19 | AR12779 | TP | 17797 | $\beta$ | −1 +0.67 |
| 59152 | −13 −22 | C1.6 C4.3 | 1.45 | AR12779 | TP | 13175 | $\beta$ | −0.4 |
| 59159 | −28 −24 −10 −9 −8 +1 +6 +21 | C1.7 C2.3 C2.8 C1.7 C2.0 C1.3 C1.8 C1.8 | 1.11 | AR12781 | TP | 19203 | $\beta - \gamma$ | +1.4 |
| 59164 | −16 | C1.3 | 1.53 | AR12781 | FN | 13570 | $\beta$ | −0.6 +0.5 |
| 59176 | −18 +11 +12 | C3.3 C1.4 C4.4 | 1.15 | AR12785 | TP | 12227 | $\beta$ | +4.3 |
| 59181 | −19 −10 +2 +11 +24 | C1.5 C1.5 C2.9 C3.1 C1.6 | 1.37 | AR12786 | TP | 15007 | $\beta - \gamma$ | +8.52 |
| 59323 | +14 | M1.1 | 1.46 | AR12816 | TP | 13297 | $\beta$ | −0.7 |
| 59360 | +10 +11 +12 | C2.9 C3.3 C3.3 | 1.13 | AR12826 | TP | 18187 | $\beta$ | −6.3 |
| 59388 | −2 | C3.4 | 1.5 | AR12833 | TP | 11616 | $\alpha$ | −0.5 |
| 59395 | +8 | C2.8 | 1.1 | AR12835 | TP | 15429 | $\beta - \gamma - \delta$ | +2.18 |
| 59456 | −24 | C6.0 | 2.04 | AR12860 | TP | 15139 | $\beta - \gamma$ | −0.7 +0.5 |
| 59565 | −29 −24 | C4.5 C6.9 | 1.11 | AR12907 | TP | 17144 | $\beta - \gamma$ | +0.68 −0.6 |
| 59572 | −8 −3 | C5.2 C5.6 | 1.49 | AR12907 | TP | 16573 | $\beta - \gamma$ | −0.47 |
| 59572 | −4 −4 +0.5 +2 | C5.4 C4.3 C4.0 C5.3 | 1.49 | AR12908 | TP | 16573 | $\beta - \gamma$ | −0.47 |
| 59572 | −7 | C7.5 | 1.62 | AR12917 | TP | 12845 | $\beta$ | −0.40 |
| 59610 | +21 | C6.0 | 1.36 | AR12940 | TP | 15509 | $\beta$ | −1.2 |
| 59610 | - | - | 1.42 | AR12939 | FP | 12642 | $\beta$ | +1.49 |
| 59610 | - | - | 1.32 | AR12936 | FP | 15049 | $\beta$ | −7.44 |
| 59619 | - | - | 1.46 | AR12941 | FP | 14493 | $\beta$ | +0.8 −0.8 |
| 59619 | - | - | 1.46 | AR12940 | FP | 13720 | $\beta$ | +0.5 −0.4 |
| 59638 | - | - | 0.93 | AR12957 | FP | 16052 | $\beta$ | −0.17 |
| 59653 | −25 +2 +12 | M1.7 M1.3 M1.4 | 1.47 | AR12965 | TP | 12799 | $\beta - \gamma$ | −1.01 +1.01 |
| 59666 | −4 +2 +2 +4 +4 +5 +10 +11 | M4 C9.7 C9.7 C6 M1 M1.1 M2.2 M1.1 | 0.88 | AR12975 | TP | 20118 | $\beta - \gamma$ | −1.2 |
| 59666 | - | - | 1.38 | AR12976 | FP | 13813 | $\beta - \delta$ | +1.83 −1.83 |
| 59688 | −27 −24 −14 −6 +26 | M1.3 M1.2 C7.2 M1.1 M1.9 | 0.98 | AR12993 | TP | 22262 | $\beta - \gamma$ | −12.4 |
| 59688 | +11 +15 | M1.6 M1.3 | 0.98 | AR12994 | TP | 22262 | $\beta - \gamma$ | −12.4 |
| 59697 | +30 | M1.2 | 1.38 | AR12994 | TP | 14983 | $\beta$ | +0.8 |
| 59708 | - | - | 1.45 | AR13004 | FP | 13312 | $\alpha$ | −0.66 |
| 59708 | +27 | X1.0 | 1.42 | AR13006 | TP | 14965 | $\beta - \delta$ | −0.89 +0.79 |
| 59708 | −15 | C7.9 | 1.58 | AR13007 | FN | 13620 | $\beta$ | +0.24 |
| 59715 | - | - | 1.47 | AR13014 | FP | 15563 | $\beta - \delta$ | −0.4 +0.4 |
| 59715 | +3 | M2.4 | 1.30 | AR13017 | TP | 15044 | $\beta - \gamma$ | 0.3 |
| 59715 | +25 | C9.8 | 1.51 | AR13010 | FN | 14696 | $\beta$ | −0.77 +0.77 |
| 59895 | −27 −24 −10 | M1.2 M1.1 M1.1 | 1.39 | AR13141 | TP | 15057 | $\beta - \gamma$ | −1.46 +0.97 |
| 59940 | −11 −2 | M2.0 M1.1 | 1.26 | AR13176 | TP | 16407 | $\beta$ | −1.5 |
| 59940 | +6 | M1.2 | 1.14 | AR13169 | TP | 16680 | $\beta - \gamma$ | −1.6 |
| 59946 | −4 | C8.7 | 1.54 | AR13176 | FN | 13004 | $\beta$ | −0.6 +0.5 |
| 59946 | - | - | 1.46 | AR13180 | FP | 15307 | $\beta$ | −0.8 +0.8 |
| 59953 | −2 +2 +15 | M2.1 C9.5 M2.7 | 1.01 | AR13181 | TP | 19494 | $\beta - \gamma - \delta$ | + 1.29 −0.96 |
| 59953 | −30 | C9.0 | 1.1 | AR13182 | TP | 17597 | $\beta - \gamma - \delta$ | +1.23 |
| 59953 | −29 −26 −25 −19 −15 −15 +7 +16 +24 +30 | C8.0 M1.2 M1.4 M1.4 C9.6 M1.1 X1.8 M1.0 C9.7 M1.0 | 0.89 | AR13184 | TP | 14794 | $\alpha$ | +3.0 |
| 59960 | +14 | C9.3 | 1.16 | AR13192 | TP | 16226 | $\beta - \gamma$ | +0.5 |
| 59960 | −17 | M4.8 | −0.51 | AR13190 | TP | 32037 | $\beta - \gamma - \delta$ | +36.99 |
| 59960 | −1 | C8.7 | −0.07 | AR13182 | TP | 24372 | $\beta$ | +1.18 |
| 59960 | −30 | M6.0 | 1.16 | AR13191 | TP | 15075 | $\beta$ | −0.5 |
| 59960 | - | - | 1.23 | AR13194 | FP | 14581 | $\beta$ | +8.17 |
| 59987 | −22 +3 | M1.4 M1.2 | 1.12 | AR13222 | TP | 16825 | $\beta$ | −1.2 |
| 59987 | −24 | M1.0 | 0.95 | AR13220 | TP | 17303 | $\beta - \gamma$ | +2.17 |
| Continued | | | | | | | | |

| Epoch | $\Delta_T$ | | Flare class | $\alpha_{T_p}$ | $AR_{ID}$ | Flag | $T_p$ | Magnetic complexity | $\rho_\%$ |
|---|---|---|---|---|---|---|---|---|---|
| [MJD] | [h] | | | | | | [K] | | [%] |
| 59987 | $-22$ $-21$ $-19$ $-5$ $-2$ $-1$ $+5$ | | M1.5 M1.1 X1.1 C9.1 M2.9 M1.4 M1.0 | 0.95 | AR13217 | TP | 17303 | $\beta - \gamma$ | +1.18 |
| 59987 | $+19$ $+29$ | | M1.0 M1.4 | 1.5 | AR13226 | TP | 13512 | $\beta - \gamma$ | $-0.59$ |
| 59987 | $-6$ | | C8.2 | 1.05 | AR13214 | TP | 19773 | $\beta$ | $+1.4$ $-0.9$ |
| 60108 | - | | - | 1.49 | AR13326 | FP | 13215 | $\alpha$ | +0.3 |
| 60108 | - | | - | 1.48 | AR13327 | FP | 15211 | $\alpha$ | $+0.5$ $-0.4$ |
| 60165 | $-16$ | | C8.7 | 1.53 | AR13394 | FN | 14301 | $\beta - \gamma$ | $+3.14$ $-0.3$ |
| 60165 | $-25$ | | M3.6 | 1.45 | AR13387 | TP | 12073 | $\beta$ | +0.2 |
| 60258 | $+29$ | | M1.2 | 1.18 | AR13477 | TP | 14705 | $\beta$ | +19.9 |
| 60259 | $+6$ | | M1.2 | 1.71 | AR13477 | FN | 12773 | $\beta$ | +13.4 |
| 60276 | $+4$ $+5$ | | M3.5 M9.8 | 1.82 | AR13500 | FN | 12655 | $\beta - \gamma$ | +13 |
| 60290 | $-5$ $+4$ $+29$ | | C9.1 C8.2 C9.3 | 1.48 | AR13511 | FN | 11207 | $\beta$ | +3.4 |
| 60290 | $-6$ $+4$ | | C9.1 C8.2 | 1.73 | AR13514 | FN | 12705 | $\beta$ | +9.9 |
| 60306 | $+21$ | | C9.7 | 1.63 | AR13533 | FN | 14355 | $\beta$ | +13.2 |

**Table 2.** AR parameters used for the analysis: the epoch of the map is expressed in Modified Julian Day [MJD]; $\Delta_T$ is the delay time in hours between the solar radio map and the detected flare (if any); Flare Class; Spectral index $\alpha_{T_p}$; NOAA AR identification number $AR_{ID}$, the index $l$ represents AR located at the limb, corresponding to a distance greater than 95% of the solar disk radius; the category (Flag) of the event respect to the solar flare occurrence (True Positive (TP), False Negative (FN), False Positive (FP)); the maximum brightness temperature corresponding to the AR detection at 18 GHz ($T_p$); the AR Magnetic Complexity; circular polarisation percentage ($\rho_\%$). Negative values are associated with RHCP, positive values with LHCP. The error associated with the $\rho_\%$ is the RMS of a background region where no sunspots have been identified on the photosphere. We obtained a mean value of $\simeq 0.05\%$. The errors on spectral index and maximum brightness temperature are typically $\sigma_\alpha < 0.1$ and $\sigma_{T_p} < 3\%$. Both were calculated in[8].

| Semester | Threshold Class |
|---|---|
| 2018 Jan-Jun | C1 |
| 2018 Jul-Dec | B4 |
| 2019 Jan-Jun | C2 |
| 2019 Jul-Dec | B2 |
| 2020 Jan-Jun | B7 |
| 2020 Jul-Dec | C1 |
| 2021 Jan-Jun | C2 |
| 2021 Jul-Dec | C4 |
| 2022 Jan-Jun | C6 |
| 2022 Jul-Dec | C7 |
| 2023 Jan-Jun | C8 |
| 2023 Jul-Dec | C8 |

**Table 3.** Thresholds for the flare intensity class, defined for each semester from 2018 to 2023.

on the photosphere. We obtained a mean value of $\simeq 0.05\%$, which is the 1/20 of the expected polarisation of the QS (1%) in K-band[31]. The errors on spectral index and maximum brightness temperature are typically $\sigma_\alpha < 0.1$ and $\sigma_{T_p} < 3\%$ respectively, as calculated in[8].

## Data availability
The datasets generated during and/or analysed during the current study are available from the PI on reasonable request and is archived in the web site https://sites.google.com/inaf.it/sundish. The flares catalogues based on GOES data are available at https://zenodo.org/records/11150339. The catalogue with the AGILE data is availa ble at https://www.ssdc.asi.it/agilesolarcat/. All other data are available in the main text or the supplementary materials.

## Acknowledgments

The Medicina Radio Telescope is funded by the Ministry of University and Research (MIUR) and is operated as National Facility by the National Institute for Astrophysics (INAF). The Sardinia Radio Telescope is funded by the Ministry of University and Research (MIUR), Italian Space Agency (ASI), and the Autonomous Region of Sardinia (RAS) and is operated as National Facility by the National Institute for Astrophysics (INAF). This research used version 3.1.3[55] of the SunPy open source software package[56].

## Author contributions

S.M., A.P., M.M., A.M., S.R., M.N.I. conceptualised and organised the project; S.M., A.P., M.M., A.M., S.R., M.N.I., E.E., G.M., F.B. analysed the scientific data; S.M., A.P., M.M., A.M., S.R., M.N.I., E.E., G.M., M.B., F.B., S.L.G, C.H, S.L, C.T, P.Z, M.M. Interpreted the results; S.M., A.P., M.M., A.M., S.R., E.E., M.N.I., G.M., A.C., R.C., G.L.D, A.L, A.M, P.M, A.M, A.N, A.O, P.O, M.P, T.P, G.P., A.S., L.S., G.S., G.V., A.Z. performed the observations and developed the instruments. All authors reviewed the manuscript

## Funding

We acknowledge the support provided by the Enhancement of the Sardinia Radio Telescope (SRT) for the study of the Universe at high radio frequencies which is financially supported by the National Operative Program (Programma Operativo Nazionale - PON) of the Italian Ministry of University and Research "Research and Innovation 2014-2020", Notice D.D. 424 of 28/02/2018 for the granting of funding aimed at strengthening research infrastructures, in implementation of the Action II.1 – Project Proposals PIR01_00010 and CIR01_00010. We recognize the support from the Italian Space Agency and the Ministry of University and Research under Contract n. 2024-5-E.0 (Space It Up). This work was supported by the Istituto Nazionale di Astrofisica (Bando per il finanziamento della Ricerca Fondamentale 2023 - IDEA-SW project).

## Declarations

### Competing interests

The authors declare no competing interests.

### Additional information

**Supplementary Information** The online version contains supplementary material available at https://doi.org/10.1038/s41598-025-27776-2.

**Correspondence** and requests for materials should be addressed to S.M.

**Reprints and permissions information** is available at www.nature.com/reprints.

**Publisher's note** Springer Nature remains neutral with regard to jurisdictional claims in published maps and institutional affiliations.